\documentclass[runningheads]{llncs}
\usepackage{graphicx}
\usepackage{amstext}
\usepackage{booktabs}
\usepackage{bbm}

\begin{document}
\bibliographystyle{unsrt}
\title{Automatic Calcium Scoring in Cardiac and Chest CT Using DenseRAUnet}
 \author{Jiechao Ma \and
 Rongguo Zhang}

 \institute{Infervision Inc.}

\maketitle              % typeset the header of the contribution
\begin{abstract}
Cardiovascular disease (CVD) is a common and strong threat to human beings, featuring high prevalence, disability and mortality. The amount of coronary artery calcification (CAC) is an effective factor for CVD risk evaluation. Conventionally, CAC is quantified using ECG-synchronized cardiac CT but rarely from general chest CT scans. However, compared with ECG-synchronized cardiac CT, chest CT is more prevalent and economical in clinical practice. To address this, we propose an automatic method based on Dense U-Net to segment coronary calcium pixels on both types of CT scans. Our contribution is two-fold. First, we propose a novel network called DenseRAUnet, which takes advantage of Dense U-net, ResNet and atrous convolutions. We prove the robustness and generalizability of our model by training it exclusively on chest CT while test on both types of CT scans.
Second, we design a loss function combining bootstrap with IoU function to balance foreground and background classes. 
%Our network employs the Residual Atrous Unit (RAU) to obtain segmentation accuracy, and utilizes an extra dense block to maximize the use of shallow image features. 
%positive and negative samples. 
DenseRAUnet is trained in a 2.5D fashion 
%in order to learn spatial information with low computational overhead.
and tested on a private dataset consisting of 144 scans. Results show an F1-score of 0.75, with 0.83 accuracy of predicting cardiovascular disease risk.

\keywords{ Calcium scoring\and deep learning\and Cardiac CT\and Chest CT\and Agatston score\and Convolutional neural network.}
\end{abstract}
\section{Introduction }

Cardiovascular disease (CVD) has become one of the most high-mortality diseases, for which the amount of coronary artery calcification acts as a strong indicator of CVD risk \cite{rumberger1999electron}. %The amount of coronary artery calcification (CAC) is a strong and independent predictor of CVD events \cite{rumberger1999electron}.
In %daily 
clinical practice, CAC is quantified by the Agatston score, using dedicated cardiac CT scans, followed by a expert who manually identify CAC lesions.
% , which can be identified and quantified in cardiac CT. Clinically, calcium scoring is performed by experts who manually identify CAC in CT image slices. But it is a time-consuming and energy-consuming process for doctors to do this work repeatedly.
% \cite{de2019direct}  
 
To assist %experts
medical professionals, previous work based on classical machine learning have attempted to design %a method which can automatically  
CAD methods for computation of CAC score. Durlak et al. \cite{durlak2017growing} applied an atlas-based feature approach in combination with a random forest classifier which is used to incorporate fuzzy spatial knowledge from offline data. Isgum et al. \cite{isgum2012automatic} 
%used both classify samples with 
employed a nearest neighbor classifier directly and a two-stage classification with nearest neighbor as well as support vector machine classifiers. There are plenty of other research can be explored \cite{kurkure2010supervised,shahzad2013vessel,wolterink2014automatic}. % Shahzad et al. \cite{shahzad2013vessel} also apply menthod based on machine learning.

In recent years, convolutional neural networks (CNNs) have exhibited great success in Computer Vision by data-driven, especially in image classification tasks. Meanwhile, fully convolutional networks (FCNs) , as the extension of CNNs, also obtained state-of-the-art performance for segmentation problems. In the context of medical image segmentation, specifically cardiac calcification segmentation, algorithms based on deep learning have shown promise. Wolterink et al. \cite{wolterink2015automatic} first attempted to apply CNNs to CAC scoring in contrast-enhanced cardiac CT, with a two-stage network structure but only one stage using deep learning. Recently, some works used two-stage deep learning structure \cite{lessmann2018automatic,wolterink2016automatic}, with the first stage identifying CAC-suspected voxels 
%likely to be CAC, filtering the majority of non-CAC-like voxels, 
and the second stage more precisely identifying CAC. Shadmi et al. \cite{shadmi2018fully} 
%is different from two-stage design which use Dense-FCN 
employed Dense-FCN, a design different from the two-stage methods, to segment the lesion directly in cardiac CT. But all the automatic CAC scoring approaches above are designed for either cardiac or chest CT only. 

More recently, multiple screening in one CT session has become a trend in clinical practice. Huang et al. \cite{huang2017densely} presented an automatic method with two CNNs that performs direct computation of CAC score in both cardiac and chest CT scans. 
%types of non-contrast non-enhanced CT. 
On the other hand, according to the work of Wolterink et al. \cite{wolterink2016automatic}, 2.5D input has a great advantage compared with 3D input in CAC scoring, as the number of parameters are greatly reduced while retaining spatial information. Both Lessmann et al. and wolterink et al. \cite{lessmann2018automatic,wolterink2016automatic} used the 2.5D ConvNets combining features from three identical 2D ConvStacks with shared weights, each processing an input patch from a different orthogonal viewing direction (axial, sagittal and coronal). To our knowledge, none have applied the efficient 2.5D FCN architecture on multiple types of non-enhanced CT.
%We present an automatic method to perform CAC segmentation both in chest CT and cardiac CT. 
 
In this work, 
%In order to improve previous methods that focused on a single type of CT, 
we propose an automatic method for CAC scoring on both ECG-synchronized cardiac and chest CT. Unlike the the methods required two cascaded networks to calculate CAC scoring, our network directly segment the calcified voxels and obtain CAC scoring. Meanwhile, we adopt a 2.5D patch input to reduce the computational overhead of 3D input. Instead of previous 2.5D methods \cite{lessmann2018automatic,wolterink2016automatic}which input patch from axial, sagittal and coronal direction, our network takes 9-channel stacks of images with corresponding 2D labels for segmentation of the corresponding center slice. 
%To our knowledge, it is the first time that our proposed 2.5D input method has been applied to the CAC scoring  in  both ECG-synchronized cardiac CT and chest CT. 
We applied our method on a private dataset composed of 44 Cardiac CT scans and 805 chest CT scans. In comparison to experts' manual annotations, our algorithm achieved competitive results.

\section{Materials and Methods}
\subsection{Data}
A dataset of 849 CT scans was collected from several medical centers in China, which consists of 805 chest CT scans and 44 cardiac CT scans. The CT scans were acquired by different CT scanners with Philips, GE and Siemens. Each CT scan contains a sequence of slices at the thin-section slice spacing (range from 1.0 to 3.0 mm). CAC lesions were manually labeled by three experienced radiologists from different centers.

We newly connected 144 CT scans as a test set, incorporating chest CT scans and cardiac CT scans from medical centers in China. And to evaluate our network performance, lesions were delineated by experienced radiologists.
% The dataset consists of 849 plain CT scans, collected from 9 hospitals to guarantee the diversity of data. The data were acquired with Philips, GE and Siemens CT scanners with layer thickness of 1-3 mm. For training, 705 scans from 3 hospitals were used, which are all lung non-gated CT. For testing, 144 scans from 7 hospitals were chosen, two of which are the same as training set.

% To train and evaluate our method, candidate CAC lesions were manually labeled by three experienced doctors from different hospitals. We calculated cardiovascular risk corresponding to Agatston Score (I:0, II:1-10, III: 11-100, IV:101-400, V:$>$400).

\subsection{Data preprocessing}
Since the connected dataset contains various sizes of chest CT scans and cardiac CT scans, we process all images as follows. First, we resize all CT images to $512 \times 512$ pixel resolution. Second, we randomly crop and then resize images to $512 \times 512$, where maximum size of the cropped image is $256 \times 256$. We continuously select nine processed slices as the input of our network, and for such an input, its label is the ground-truth label of its middle slice. We also process all ground-truth labels to alter the pixel label when its corresponding CT value lower than 130HU.

%  For data preprocessing, we resized the image to the standard resolution for normalization and then randomly cropped followed by resizing to 512x512 pixels. In the channel number dimension, the first four layers and the last four layers of each graph are merged in order (filled with zero when the boundary is exceeded), forming a 9-channel 2.5D array for the model to predict the calcification score at the central level. We also balanced positive and negative samples.
 
\subsection{DenseRAUnet for segmentation}
\begin{figure}
\centering
\includegraphics[width=\textwidth]{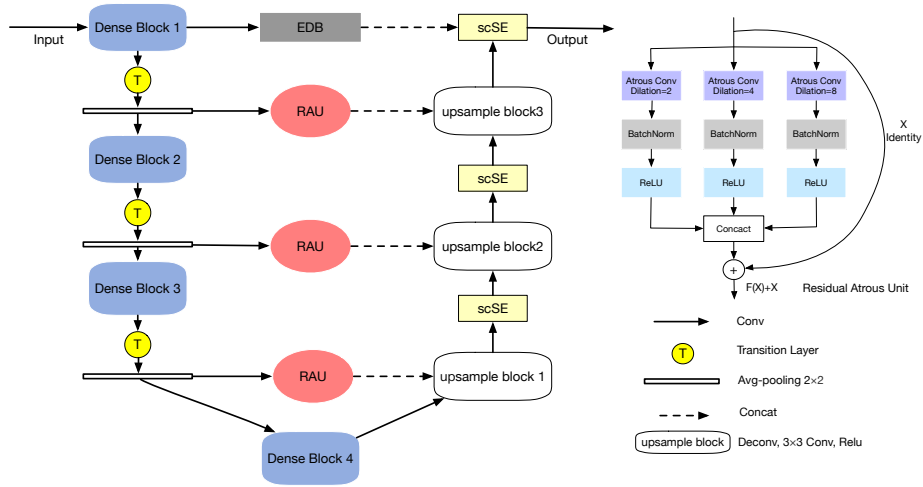}
\caption{The overall structure of DenseUnet and details in Residual Atrous Unit } \label{fig1}
\end{figure}

We proposed a novel FCN architecture based on dense U-Net for calcification segmentation, called DenseRAUnet. The network consists of two main components: (1) a basic network for feature extraction, and (2) three task-specific sub-network structures, incorporating \emph{Residual Atrous Unit (RAU)}, \emph{scSE block} and \emph{Extra Dense Block (EDB)}. Fig.~\ref{fig1} depicts our proposed DenseRAUnet.

%基础网络描述
The basic network is an encoder-decoder architecture, similar to dense U-Net. We adopt a backbone network (DenseNet-121) as the encoder sub-network. The decoder sub-network consists of three decoder modules. Each decoder module is an upsampling block followed by a scSE block, where upsampling block contains a deconvolution layer and two convolution layers, which followed by a Batch Normalization (BN) layer and an activation function called ReLU.
% 
% Our network is an encoder-decoder structure. The encoding path contains 4 stages, in which we cascade a dense block and a Residual Atrous Unit (RAU). In the decoding sequence, we add scSE block between upsample blocks. 
%  Our proposed method for automatic CAC scoring use U-net[10] as basic network structure. In the encoding path, we adopt denselayer which combined with denseblock and our self-designed  Residual Atrous Unit (RAU) and an Extra Dense Block (EDB) for better transmission. In the decoder, We used the scSE to better restore details. The overall structure of the network is shown in Fig.2.

% \begin{figure}
% \includegraphics[width=5cm,height=5cm]{fig/raun.eps}
% \centering
% \caption{The RAU unit} \label{fig1}
% \end{figure}

\noindent\textbf{Residual Atrous Unit.} Accurately segmenting various sizes of calcified areas may require different combinations of local and global information. So we consider that a simple skip connection is not enough for the complex segmentaiton task. Inspired by ASPP \cite{chen2018deeplab} and embed the idea of Inception \cite{szegedy2015going}, we further design a lateral connection called Residual Atrous Unit (RAU). Such a module is a residual block, and is used to capture multi-scale information by combining several convolutional layers with different dilation rates in parallel. As shown in Fig.～\ref{fig1}, we use a concatenation of three $3 \times 3$ dilated convolution layers with dilation rates are 2, 4, and 8 in each RAU.

%\begin{figure}
%\includegraphics[width=6cm,height=6cm]{fig/raun.eps}
%\centering
%\caption{Residual Atrous Unit} \label{fig2}
%\end{figure}
% We designed RAU structure inspired by residual connection\cite{he2016deep} and atrous convolution\cite{chen2017rethinking}. The unit consists of three 3*3 dilated convolution layers with dilation rate = [2,4,8], each atrous is concat with others, as after pre-processing, some of the lesions in the input images become larger and require larger receptive fields. The purpose of the unit is to quickly expand the receptive field without deepening the network and to process images with different magnification. Residual connection ensure effective gradient propagation and make the model learn parameters better.

\noindent\textbf{scSE block.} To take full advantage of local and global information, we added scSE block in the decoder sub-network, which is introduced in \cite{roy2018concurrent} for recalibrating the feature maps separately along channel and space.
% We added scSE block\cite{roy2018concurrent}as attention mechanism in decoder because the scSE block can learn important features from both channel and spatial information in the network without adding parameters. 

\noindent\textbf{Extra Dense Block.} In order not to waste the image features extracted from input images, we insert an Extra Dense Block (EDB) in the first skip connection. Such a block could make more accurate use of shallow information, which do not represent input image in a high dimensional space, via adding more nonlinear into the first long connection.
% The composition of the EDB is the same as Dense block.
% To restore details better in decoder, we designed an extra block following stage1, as upsampling block needs to concact with downsampling block to restore details in U-net structure. However, the shallow information (such as stage1)  contribute little to final result. Therefore, we added an extra block to stage1 to make shallow information deeper and obtain more abstract features. 
% \begin{figure}
% \includegraphics[width=\textwidth]{fig/newall.eps}
% \caption{The overall structure of DenseUnet} \label{fig2}
% \end{figure}
\subsection{Loss Function}% One of the main problems in medical image processing is the imbalance of positive and negative samples, a problem also present in calcification score calculation. To solve the segmentation of calcification, we designed a combination of two loss functions with emphasis on different aspects:
Inter-class imbalances are common problems when using deep learning methods for image segmentation, and even more in medical image segmentation. To solve it, we propose a new loss function, the combination of Bootstrap Loss and IoU Loss:
\begin{equation}
Loss = Bootstrap\ Loss + IoU\ Loss
\end{equation}
\noindent\textbf{Bootstrap Loss.} When we train a FCN, though images were cropped, there may be thousands of labeled pixels to predict. However, many of them may be easily distinguishable, and continuing to learn from these pixels does not improve model performance. In the context of medical image segmentation, most of such pixels are marked as background. For this reason, we design a weighted bootstrap loss, which not only forces network to focus on hard pixels but also balances positive and negative pixels during training.

Suppose there is only one processed image per mini-batch and there are a total $N$ pixels to predict. There are only two categories $c_{j}$ in the label space. Let $y_{i}$ denotes the ground-truth label of pixel $x_{i}$, and $p_{i,j}$ denotes the predicted probability that pixel $x_{i}$ belongs to the category $c_{j}$. Then, the loss function could be defined as:
% Bootstrap loss mainly solves two difficulties of this problem: 1). the extreme imbalance of positive and negative samples of training data. This is more of a problem in lung CT than in cardiac CT, even at the layer of positive label, calcification is found only in small section of the coronary artery of the heart, the rest is mostly background. 2). the loss of cross entropy can not converge, because the loss contribution of a large number of simple negative samples masked the contributions of difficult negative samples and positive samples, so the model cannot consistently learn in the right direction. Using this loss function enables the model to converge quickly, the loss function is defined below:

\begin{equation}
% Bootstrap\ Loss =  E(\sum_{t}[ -\alpha y_tlogP_t- \beta( 1-y_t) \cdot mask \cdot log(1-P_t)])
l = - \left\{\alpha \frac{\sum\limits_{i \in N, j=0}log p_{i,j}1\left\{p_{i,j} < t\ and\ y_{i}=j\right\}}{\sum\limits_{i \in N, j=0}1\left\{p_{i,j} < t\ and\ y_{i}=j\right\}} + \beta \frac{\sum\limits_{i \in N, j=1}log p_{i,j}1\left\{y_{i}=j\right\}}{\sum\limits_{i \in N, j=1}1\left\{y_{i}=j\right\}}\right\}
\end{equation}
where $t$ is a threshold. Here $1\left\{\cdot\right\}$ is equal to one when the condition in parentheses, and otherwise is zero. In other words, we focus all positive pixels and drop negative pixels when they are too easy for the current model, i.e. their predicted probability greater than $t$. In practice, we hope that positive and negative pixels are balanced, hence we add $\alpha$ and $\beta$ as trade-off coefficients.
% where $y_t$ is ground-truth label for each pixel,$P_t$ is the prediction label,$\alpha,\beta$ are parameters of our custom balanced positive and negative samples. We design mask as:$(1-P_t)<0.9$ to filter simple negative samples. E(.) represents average value.
% During the training period, to achieve better convergence and accurate prediction,we manually adjusted the values of $\alpha$ and $\beta$. Finally, the best results can be obtained when $\alpha$ = 8 and $\beta$ = 1.

\noindent\textbf{IoU Loss.} Bootstrap loss is similar to cross entropy loss, focusing more on its own predictions of pixels and ignoring the relationship between adjacent ones. To better obtain the boundary of lesion, we add IoU in the loss function using such a relationship. Suppose there are $N$ pixels to predict. To ensure that losses are on the same magnitude, we use the following exponential form of IoU:
% We consider that bootstrap loss only takes into account the loss of a single pixel, but does not represent the relationship between pixels and pixels, because in calcification calculation, no less than 3 connected calcification pixels can be defined as a calcification lesion, which shows that there is influence between pixels, so we added IoU loss to make up for this deficiency:

\begin{equation}
% IoU\ Loss=-ln \frac{Pred\cap Gt} {Pred\cup Gt}
iou\ loss = -ln \frac{\sum\limits_{i \in N}p_{i}g_{i}}{\sum\limits_{i \in N}p_{i} + \sum\limits_{i \in N}g_{i} - \sum\limits_{i \in N}p_{i}g_{i}}
\end{equation}
where $p_{i}$ is the predicted probability of pixel $x_{i}$, $g_{i}$ is the ground-truth label of pixel $x_{i}$.
% where Pred represents area of prediction, Gt represents area of ground truth.

\subsection{Post-processing}
The final segmentation result of the network is obtained by a predefined threshold (here set to 0.5), and each lesion segmented by the network is considered a calcification candidate. Then each candidate is classified as CAC by thresholding with 130 HU and performing connected-components analysis. Since the CT slice thickness is mostly 1mm, calculation of the final Agatston score for the whole volume is done by the following corrected formula:
% The softmax layer outputs the probability map of candidate pixels which belongs to a CAC lesion. Firstly, we selected pixels with probability greater than the threshold 0.5, and original map corresponding to a CT value greater than 130 HU, then find the area connected region in each layer larger than 1 mm$^2$. The Agatston calcification integral of the whole layer is calculated according to the Hu peak of each connected region. As Agatston integral is defined with 3mm thickness, and most of our lung CT data layers are about 1mm thick, so we need to multiply the layer-spacing normalization factor with the following formula:

\begin{equation}
Agatston\ Score =\sum_{i} \sum_{n} f_{i, n} A_{i, n} \frac{\Delta S}{3}
\end{equation}

where $i$ is the $i$th CT slice of a CT volume, $n$ is the $n$th selected lesion, $f$ is the weighted intensity, $A$ is the lesion area, and $\Delta S$ is the slice spacing (mm).

\section{Experiments and Results} 
\noindent\textbf{Evaluation Metric.} We evaluate the pixel-level segmentation performance of the network by F1 score:
\begin{equation}
F1 = 2 \cdot \frac{Precision \cdot Recall}{Precision + Recall}
\end{equation}
We also define \emph{CAC rate} denotes the proportion of patients who was correctly predicting the CVD risk level without post-processing, and \emph{CAC filter Rate} represents the proportion of patients with post-processing.
% We also define CAC rate and CAC filter Rate that represent the percentage of patients in the total number of patients whose CVD risk level is correctly predicted by the model and predicted after post-processing respectively.

\noindent\textbf{Implementation details.} The experiments conducted were all trained from scratch and  initialized by the Gauss method. During training, we collected one processed image as a mini-batch for each iteration and trained for 25 epochs. To optimize these experiments with fast convergence, we employed the SGD optimizer with  momentum of 0.9. The initial learning rate is 0.001 and is reduced by 0.99 times per 2000 iterations. The parameters in the loss function are experimentally set as $t = 0.9$, $\alpha = 8$ and $\beta = 1$. We implemented all the experiments via the deep learning toolki MXNet and trained on a GTX 1080 (NVIDIA) GPU.

% We trained both Dense U-Net using bootstrap loss (baseline) and our proposed model. During training, for the decoding path, the depth of each layer of dense block is 3,6,12,8, and is half of the corresponding depth of the DenseNet-121, and the growth rate is 32. In the process of model convergence, the proportion of positive samples is continuously increased, 3 to 4 adjustments were made before the final proportion of positive and negative samples is obtained.

% We use MXNET to build up deep convolutional architecture. Our model is trained about 25 epoches with batch image of 1 for device with 1 GPU . And the network was trained using the SGD optimizer with multistep learning rate without pre-trained parameters. The initial learning rate is set to 0.001 and than gradually reduced  through multiplying  by 0.99  for each 2000 batch. 

\noindent\textbf{Ablation experiments.} We use ``Dense U-net \& Bootstrap Loss'' as the baseline for all experiments. To evaluate the effectiveness of various structures in our method, we conducted ablation experiments. First, using the bootstrap loss, we compare the role of three modules in the network. Second, we studied the effect of two loss functions through trained our network.
% In ablation experiments, we tested different combinations of the blocks in our model with internal dataset to assess the positive effect of each block. All the contrast experiments were carried out on the basis of bootstrap loss to converge as soon as possible. As shown in Table1, The overall effect of all blocks of the model increased F1-Score by 0.1.

% \begin{table}[]
% \begin{center}
% \caption{F1-Score}\label{tab1}
% % \scriptsize
% \begin{tabular}{@{}ccccccc@{}}
% \toprule
% & bootstrap & RAU  & extra dense block & iou & scSE & F1-score \\
% \cmidrule
% Dense U-Net & $\surd$& & & & & 0.65\\
% & $\surd$& $\surd$&&& & 0.68\\
% & $\surd$& $\surd$& $\surd$&&& 0.69\\
% & $\surd$& $\surd$& $\surd$&$\surd$&& 0.71\\
% & $\surd$& $\surd$& $\surd$&$\surd$& $\surd$ & 0.75\\
% Dense U-Net(simple)&$\surd$&&&&$\surd$&  0.64\\
% \bottomrule
% \end{tabular}
% \end{center}
% \end{table}

\begin{table}
\begin{center}
\caption{Comprasion of performance of the basic network using different tricks}
\label{tab1}
\scriptsize
\begin{tabular}{cccccccc}
\toprule
Basic network & Bootstrap & RAU & EDB & scSE & IoU & F1-Score & \\
\midrule
Dense U-Net & $\surd$ &         &         &         &         & 0.65 & \\
            & $\surd$ & $\surd$ &         &         &         & 0.68 & \\
            & $\surd$ & $\surd$ & $\surd$ &         &         & 0.69 & \\
            & $\surd$ & $\surd$ & $\surd$ & $\surd$ &         & 0.71 & \\
            & $\surd$ & $\surd$ & $\surd$ & $\surd$ & $\surd$ & \textbf{0.75} & Ours\\
% \midrule
% Dense U-Net(simple) &$\surd$  &  &  &  & $\surd$ &  0.64\\
\bottomrule
\end{tabular}
\end{center}
\end{table}

% Accuracy of cardiovascular risk rating for patients
\begin{table}[]
\begin{center}
\caption{ Qualitative results of CAC rate and CAC filter rate for patients, Patients represents the total number of patients, CAC No. and CAC fliter No. represents the number of patients were predicted correctly by model and post-processing respectively.}
\label{tab1}
\scriptsize
\begin{tabular}{lccccc}
\toprule
Tricks &CAC No. & CAC filter No. & Patients & CAC Rate & CAC filter Rate \\
\midrule
Dense U-net & {\itshape} 101& {\itshape} 99 & {\itshape} 144 & {\itshape} 0.70 & {\itshape}0.69\\

Dense U-net+RAU & {\itshape} 104& {\itshape} 111 & {\itshape} 144 & {\itshape} 0.72 & {\itshape}0.77\\

Dense U-net+RAU+EDB  & {\itshape} 109& {\itshape} 115 & {\itshape} 144 & {\itshape} 0.76 & {\itshape}0.80\\

Dense U-net+RAU+EDB+scSE & {\itshape} 109& {\itshape} 117 & {\itshape} 144 & {\itshape} 0.76 & {\itshape}0.81\\

% Dense U-net(simple)+RAU+atten & {\itshape} 95& {\itshape}96& {\itshape}144& {\itshape}0.66& {\itshape}0.67\\

Our proposed method & {\itshape} \textbf{113} & {\itshape} \textbf{120} & {\itshape} 144 & {\itshape} \textbf{0.78} & {\itshape} \textbf{0.83}\\
\bottomrule
\end{tabular}
\end{center}
\end{table}

Table 1 lists the F1 scores of Dense U-Net using different tricks, Table 2 indicates the performance of corresponding network architectures in Table 1 on CAC. It is shown that all the tricks provide increase in F1 score and CAC in comparison to the baseline. We further observe that adding RAU in the network achieves more significant improvement for CAC segmentation. Comparing the results across Table 1, our method yields the best performance. From Table 2 we can also conclude that post-processing by the definition of CAC score is essential. 
%Fig.~\ref{fig3} describes qualitative results on chest CT and cardiac CT.
% Our model also validated the accuracy of cardiovascular risk rating for patients as shown in Table 2, where Patients represents the total number of patients, and CAC No. represents the number of patients diagnosed correctly by the model. CAC filter No. represents the number of patients diagnosed correctly after post-processing. Among them, the indicator to determine whether the prediction is correct is based on the level of cardiovascular risk corresponding to Agatston Score. The prediction is considered correct if the calculated calcification score corresponds to the same level of cardiovascular risk as the actual level. Obviously, compared with CAC rate, the CAC filter rate are improved, it is contribute to the post-processing screening out some false positive samples. 

\section{Conclusion}
This paper proposed an algorithm based on deep learning. Our method consists of two core elements: (1) a novel fully convolutional network, DenseRAUnet, and (2) a loss function combined bootstrap loss and IoU. We trained our network in a 2.5D-patch fashion to reduce input parameters while preserving spatial information. While trained solely on chest CT, our model achieved competitive and robust performance on both chest CT and cardiac CT which has significant higher resolution and lower spacing compared to training data, thanks to the power of residual atrous unit that enlarges the receptive field with downward compatibility. 
%Our method achieve stable performance and good segmentation results. 
%We aim our method to other medical image analysis with similar challenges in the future work.
We aim to futher explore and extend our method to other medical image analysis challenges in future work.
% We proposed a novel network for CAC, with competitive results on both chest CT and cardiac CT. 2.5D input allows the model not only to contain spatial information, but also to reduce parameters greatly compared with 3D input. At the same time, we deepen the stage1 and combine the attention mechanism to recover the spatial resolution better. To the best of our knowledge, this is the first time bootsrap loss is used in the calcification score calculation, and an improved result has been obtained by adding IoU Loss. We expect that cardiac CT scans and abnormal samples will be added to the training set in the future to make the network prediction more accurate.
\begin{figure}
\centering
\includegraphics[width=12.0cm,height=3.5cm]{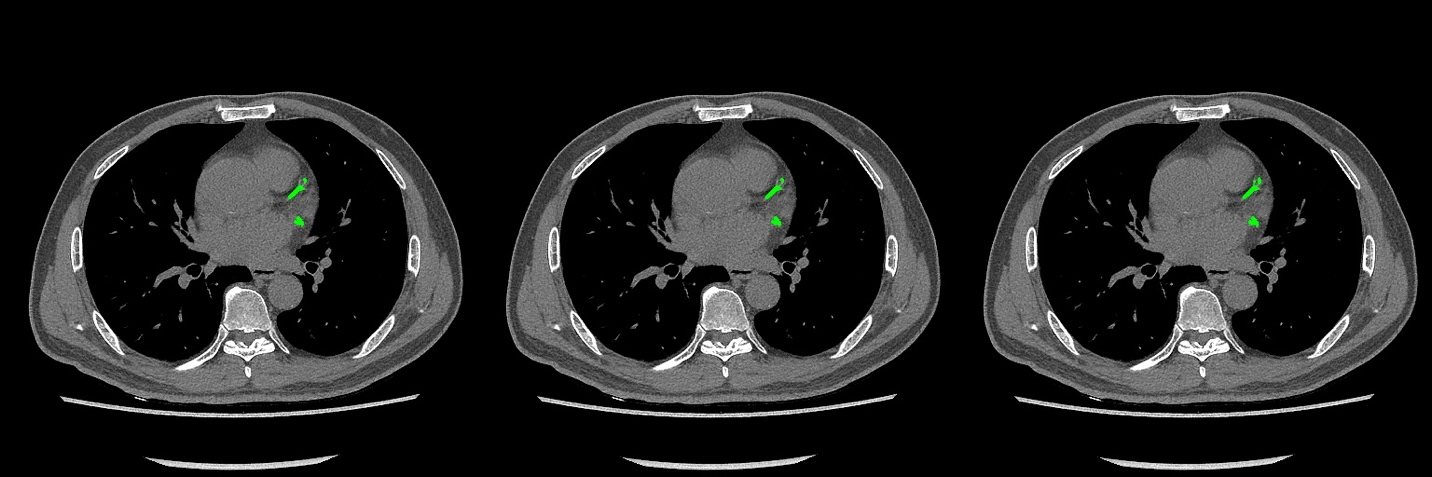}
\includegraphics[width=12.0cm,height=3.5cm]{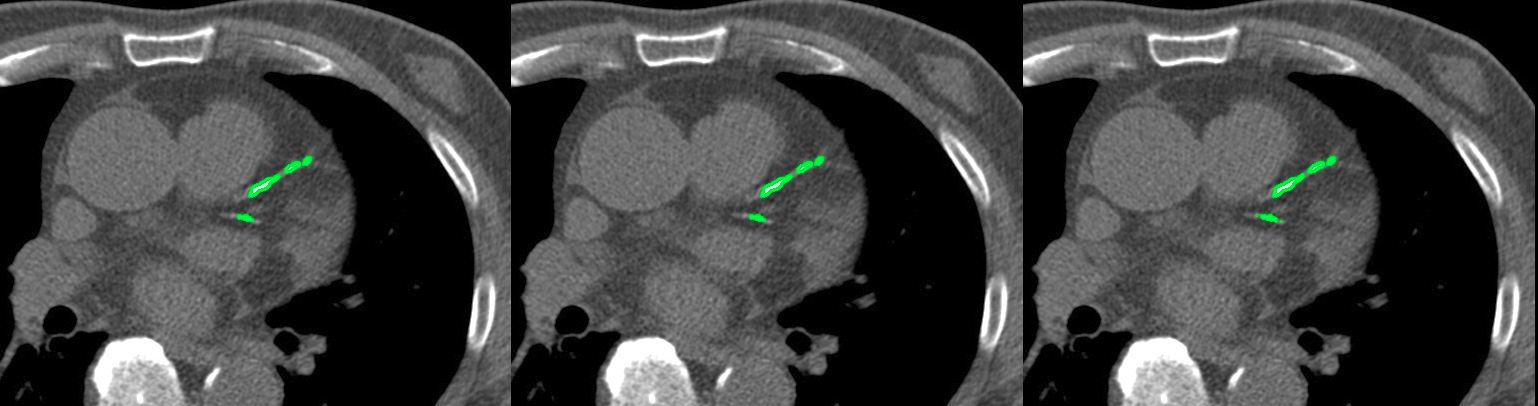}
\caption{Segmentation results of chest CT (top) and cardiac CT (bottom). From left to right: the segmentation result of our model without post-processing, the result of with post-processing, ground truth.} \label{fig3}
\end{figure}
% The segmentation result of chest CT(top) and cardiac CT(bottom).in each picture, the left side is the model's prediction, the middle is the post-processing prediction, and the right is ground truth

\bibliography{refs}
\end{document}